\documentclass[reprint, aps, nofootinbib, prd,  floatfix]{revtex4-2}
\usepackage{amsmath}
\usepackage{amssymb}
\usepackage{tensor}
\usepackage{physics}
\usepackage{xfrac}
\usepackage{color}

\usepackage{standalone}

\usepackage[autostyle]{csquotes}
\MakeOuterQuote{"}

\usepackage{graphicx} 
\setkeys{Gin}{draft=false} % graphicx inherets draft mode from \documentclass and doesn't display figures (only showing placeholders). This overrides the option so that graphicx displays figures in draft mode
\graphicspath{{figures/}}

\usepackage[final, hidelinks]{hyperref}
\hypersetup{linktoc=all}
\usepackage{cleveref}

\newcommand\scri{\mathcal{I}}
\newcommand\half{\frac{1}{2}}

\renewcommand\d{\ensuremath{\partial}}
\newcommand\Jv{J_c}
\newcommand\Jpv{J^{\prime}_c}

\newcommand\volelement{\sqrt{\abs{g}}}

\usepackage{ifdraft}
\ifdraft{
    \usepackage{setspace}
    \doublespacing
}{}
\usepackage[normalem]{ulem}

\begin{document}
\title{Evaporation of regular black holes in 2D dilaton gravity}

\author{Jonathan Barenboim}
\email{jonathan\_barenboim@sfu.ca}
\affiliation{Department of Physics, Simon Fraser University, Burnaby, BC, V5A 1S6, Canada}
\author{Andrei V.\ Frolov}
\email{frolov@sfu.ca}
\affiliation{Department of Physics, Simon Fraser University, Burnaby, BC, V5A 1S6, Canada}
\author{Gabor Kunstatter}
\email{g.kunstatter@uwinnipeg.ca}
\affiliation{Physics Department, University of Winnipeg, Winnipeg, Manitoba, R3B 2E9, Canada}
\affiliation{Department of Physics, Simon Fraser University, Burnaby, BC, V5A 1S6, Canada}

\begin{abstract}
We present a general class of non-singular black holes in semi-classical, two--dimensional dilaton gravity, with a focus on a Bardeen-like model. The equations of motion for an evaporating black hole including backreaction are solved numerically. The apparent horizons evaporate smoothly in finite time to form a compact trapped region. Backreaction effects lead to the formation of additional trapped and anti-trapped regions after the primary black hole becomes un-trapped. Numerical simulations of microscopic black holes yield a final spacetime that is free of apparent horizons and Cauchy horizons. This would imply that the evaporation of regular black holes is a unitary process.
\end{abstract} 

\maketitle

\ifdraft{\tableofcontents}

\section{Introduction}
The prediction that black holes emit radiation and the possibility that the formation and evaporation of black holes is not unitary points to an incompatibility between quantum mechanics and general relativity \cite{HawkingBreakdownPredictability1976}. Several resolutions to the so-called "black hole information loss problem" have been proposed, but the unitarity of the process remains a subject of debate (for reviews of the information loss problem and proposed solutions, see \cite{MarolfBlackHole2017,*PolchinskiBlackHole2017,*HarlowJerusalemLectures2016} and references therein).

One potential resolution relates the issue of information loss to the black hole singularity \cite{HaywardDisinformationProblem2005,AshtekarBojowaldBlackHole2005} - the singularity represents a region where the equations of motion break down and predictability is lost, but in its absence information should, in principle, be preserved, and if the region becomes untrapped all information would be accessible to outside observers. 

Complete evaporation of non-singular black holes has been studied in, for example \cite{HaywardFormationEvaporation2006,FrolovInformationLoss2014}. In such models the trapped region is compact and apparent horizons vanish in finite time. Information inside the black hole would therefore be free to escape once evaporation is completed.

However, the potential of regular black holes to resolve the black hole information loss problem has been challenged on the basis that most classical non-singular black hole models contain a Cauchy horizon\footnote{One notable exception is the single horizon loop quantum gravity inspired black hole \cite{PeltolaKunstatterCompleteSinglehorizon2009,*PeltolaKunstatterEffectivePolymer2009,ModestoLoopQuantum2006,*ModestoBlackHole2008,*ModestoSemiclassicalLoop2010}}, which introduces new pathologies. Such spacetimes are highly unstable to small perturbations, leading to mass inflation and new singularities at the Cauchy horizon, threatening the self-consistency of the model \cite{PoissonIsraelInnerhorizonInstability1989, Carballo-RubioEtAlViabilityRegular2018}. Recent investigations into the problem of mass inflation in the context of dynamic non-singular black holes have suggested that Hawking radiation ameliorates the exponential mass buildup near the Cauchy horizon \cite{BonannoEtAlRegularEvaporating2023} at late times. Nonetheless, mass inflation at early times may still be significant enough to threaten the viability of such models \cite{Carballo-RubioEtAlMassInflation2024a}. 

However, many models assume thermal and quasi-static behavior for Hawking radiation, which implies that the black hole takes infinite time to evaporate \cite{Carballo-RubioEtAlViabilityRegular2018}, or at least that the horizons evolve slowly. Complications associated with Cauchy horizons could in principle be avoided if the black hole evaporates fast enough and the final spacetime is free of horizons. Therefore it is important to understand not only the endpoint of evaporation and the disappearance of the trapped region, but also the complete structure of the spacetime in order to determine whether these models offer a viable solution to the black hole information loss problem.

In the absence of a complete theory of quantum gravity, two-dimensional dilaton gravity (DG) serves as a useful toy model for studying black hole dynamics. These models can reproduce qualitative features of higher dimensional gravity and describe Hawking radiation, including the backreaction on the spacetime, while remaining more tractable. 

Investigations of non-singular black holes in DG have found regular spacetimes with black holes evaporating in finite time \cite{LoweOLoughlinNonsingularBlack1993,DibaLoweNearextremalBlack2002,CadoniEtAlEvaporationInformation2023}. However, these studies suggest that the black hole settles into an extremal state with an eternal horizon. This causally disconnects outside observers from the interior of the black hole, and so information, though not globally destroyed, would be inaccessible outside the black hole. 

In this paper, we present a general model of 2D dilaton gravity that is capable of describing a broad range of spacetimes, including non-singular black holes\footnote{Some of the results were first presented in an earlier letter \cite{BarenboimEtAlNoDrama2024}.}. As a concrete example we focus on a Bardeen-like model. 
Our key results obtained for microscopic black holes are that in the semi-classical Bardeen model, the trapped region vanishes in finite time, leaving a regular spacetime that is free of singularities or global horizons. Additionally, small black holes do not appear to be subject to the instabilities associated with classical two-horizon black holes.

This paper is organized as follows: in \cref{sec:classical_dilaton_gravity} we introduce the general action and equations of motion. In \cref{sec:black_hole_models} we construct non-singular black hole solutions and discuss their features in the classical theory. \Cref{sec:hawking_radiation,sec:semi_classical_models} describe how Hawking radiation is added and how the black hole solutions behave in the new semi-classical theory. Finally, numerical results are discussed in \cref{sec:results_schwa,sec:results_bardeen}. Details of the numerical methods are laid out in the Appendix.

\section{Classical Dilaton Gravity}\label{sec:classical_dilaton_gravity}
We begin with a generic action
\begin{equation}\label{eq:general_action}
\begin{aligned}
    S =& \frac{1}{G} \int \volelement \left[ \tilde{\Phi}(\tilde{r}) R + K(\tilde{r}) (\nabla \tilde{r})^2 + l_p^{-2} V(\tilde{r}) \right] d^2 x \\
    &- \int \volelement B(r) (\nabla f)^2 d^2 x
\end{aligned}
\end{equation}
where $\tilde{\Phi}, K, V, B$ are functions of a scalar dilaton field $\tilde{r}$, $R$ is the Ricci curvature scalar, $f$ is a massless scalar field, $G$ is the (two-dimensional) gravitational constant, and $l_p$ is a length scale, normally but not necessarily associated with the Planck length, that is required to give the actions the correct dimensions. This is the most general action containing at most second derivatives of the metric and dilaton.

Two of these functions in (\ref{eq:general_action}) are redundant in the description and can be removed through a Weyl transformation and redefinition of the dilaton field $\tilde{r}$ \cite{MartinezEtAlExactDirac1994, GrumillerEtAlDilatonGravity2002}. In particular, we can choose $K(\tilde{r}) = V(\tilde{r}) = \tilde{\Phi}^{\prime\prime}(\tilde{r})$, so the gravitational part of the action becomes
\begin{equation}
    S_G = \frac{1}{G} \int \volelement \left[ \tilde{\Phi}(\tilde{r}) R + \tilde{\Phi}^{\prime\prime}(\tilde{r}) (\nabla \tilde{r})^2 + l_p^{-2} \tilde{\Phi}^{\prime\prime}(\tilde{r}) \right] d^2 x.
\end{equation}
As will be discussed later, the dilaton field can be related to an areal radius. To make this more apparent, we redefine the field as $r=l_p \tilde{r}, \Phi(r) = l_p^2 \tilde{\Phi}(\tilde{r})$ so that it takes on the dimensions of length. The action is then
\begin{equation}\label{eq:action}
    S_G = \frac{1}{G l_p^2} \int \volelement \left[ \Phi(r) R + \Phi^{\prime\prime}(r) (\nabla r)^2 + \Phi^{\prime\prime}(r) \right] d^2 x
\end{equation}
and for simplicity we henceforth set $l_p = 1$.

This frame is chosen for its connection to higher-dimensional gravity -- with $\Phi(r) = r^2$, \cref{eq:action} is the Einstein-Hilbert action for a 4D spherically symmetric spacetime -- and so that the static vacuum solution for general $\Phi$ can be parameterized in a form adapted to asymptotically flat spacetimes. The equations of motion derived from \cref{eq:action} are
\begin{subequations} \label{eq:EoMs_covariant_form}
\begin{equation}
    -J \left(\nabla_\mu \nabla_\nu r - g_{\mu \nu} \nabla^2 r \right) + \frac{J^\prime}{2} g_{\mu\nu} (\nabla r)^2 - \frac{J^\prime}{2} g_{\mu \nu} = G T_{\mu\nu} \label{eq:metric_eom}
\end{equation}
\begin{equation}
    JR - J^{\prime\prime} (\nabla r)^2 - 2 J^\prime \, \nabla^2 r + J^{\prime \prime} = B^\prime (\nabla f)^2 \label{eq:dilaton_eom}
\end{equation}
\begin{equation}
    g^{\mu \nu} \nabla_\mu (B \nabla_\nu f) = 0\label{eq:matter_eom}
\end{equation}
\begin{equation}
    T_{\mu\nu} =  B \left( \nabla_\mu f \, \nabla_\nu f - \frac{1}{2} g_{\mu\nu} (\nabla f)^2 \label{eq:stresstensor_eom}\right)
\end{equation}
\end{subequations}
where $J(r) = \Phi^\prime(r)$. Throughout this paper the argument will typically be dropped from functions of the dilaton field, and a prime will indicate derivatives with respect to $r$.

In the absence of matter, $f = 0$, there is a unique vacuum solution up to a parameter $M$, the mass of the system \cite{Louis-MartinezKunstatterBirkhoffTheorem1994}. By choosing $r$ as the radial coordinate the solution can be expressed as
\begin{equation} \label{eq:vac_metric_tr_coords}
    ds^2 = -\left(1 - \frac{2M}{J}\right) dt^2 + \left(1 - \frac{2M}{J}\right)^{-1} dr^2.
\end{equation}

Modeling dynamical systems, including gravitational collapse, requires coupling to matter. We will consider only minimal coupling\footnote{Note that a matter field minimally coupled in four dimensions will couple to the dilaton in the dimensionally reduced theory with $B(r) = r^2$.}, i.e. $B(r) = \text{constant}$ (the value of which is not important here). The matter field then satisfies the wave equation on flat space, 
\begin{equation}
    \nabla^2 f = 0 \Leftrightarrow \nabla^2_{(\eta)} f = 0,
\end{equation}
which implies matter distributions of the form 
\begin{equation}
    f(u, v) = f_u(u) + f_v(v),
\end{equation}
where $u, v$ are arbitrary double null coordinates. To model a black hole formed by the collapse of an infinitely thin null shell two different vacuum metrics (\cref{eq:vac_metric_tr_coords}), an interior solution with $M=0$ and an exterior solution with $M>0$, are joined along the surface $v = v_0$ (see \cref{fig:ex_penrose}). The resulting stress-energy tensor is a shock wave traveling to the left,
\begin{equation} 
\begin{aligned}
    T_{uu} &= B (\d_u f \, \d_u f) = 0, \\
    T_{vv} &= B (\d_v f \, \d_v f) = M \delta(v - v_0).
\end{aligned}
\end{equation}
$T_{uv}$ is identically 0 for any massless $f$ in the classical theory as a consequence of conformal invariance, which implies that the stress-energy tensor is traceless. In this setup \cref{eq:metric_eom,eq:dilaton_eom} decouple from \cref{eq:matter_eom,eq:stresstensor_eom} and the system for the exterior region $v > v_0$ reduces to the equations
\begin{subequations}\label{eq:EoMs_covariant_form_shell}
\begin{equation}
    -J \left(\nabla_\mu \nabla_\nu r - g_{\mu \nu} \nabla^2 r \right) + \frac{J^\prime}{2} g_{\mu\nu} (\nabla r)^2 - \frac{J^\prime}{2} g_{\mu \nu} = 0
\end{equation}
\begin{equation}
    JR - J^{\prime\prime} (\nabla r)^2 - 2 J^\prime \, \nabla^2 r + J^{\prime \prime} = 0.
\end{equation}
\end{subequations} 

\begin{figure}[hpbt]
    % \includestandalone[mode=buildnew, width=0.7\linewidth]{penrose_classical_single_horizon}
    \includegraphics[width=0.7\linewidth]{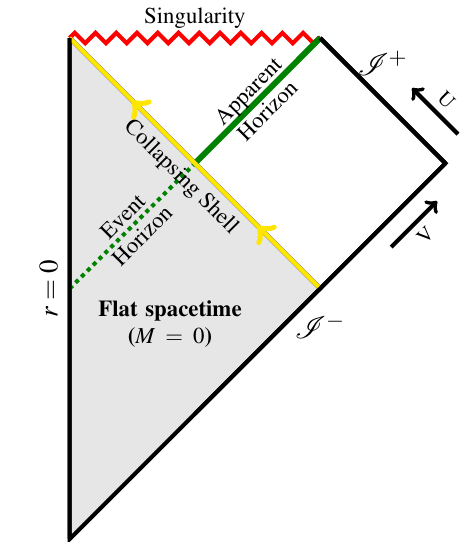}
    \caption{Example Penrose diagram for a classical, single horizon,  singular black hole formed by the collapse of a null thin shell.}
    \label{fig:ex_penrose}
\end{figure}

Two-dimensional theories are most convenient to study in the conformal gauge, $g_{\mu\nu} = e^{2\rho} \eta_{\mu \nu}$, and in particular we use a null coordinate system which matches the flat interior solution in Minkowski coordinates along the shell. These are given by first introducing the asymptotically Minkowski coordinates
\begin{equation} \label{eq:vac_metric_asymp_coords}
    ds^2 = \left(1 - \frac{2M}{J}\right) d\bar{u} \, dv
\end{equation}
with $v = t + x, \bar{u} = t-x, dr = \left(1 - \frac{2M}{J}\right) dx$, and then making the transformation 
\begin{equation}\label{eq:u_coord_transform}
    d\bar{u} = \left(1 - \frac{2M}{J(r(u, v_0))}\right)^{-1} du.
\end{equation}
The exterior metric becomes
\begin{equation} \label{eq:vac_metric_null_coords}
    ds^2 = -e^{2\rho} du \, dv = -\left(1 - \frac{2M}{J}\right) \left(1 - \frac{2M}{J_0}\right)^{-1} du \, dv
\end{equation}
where $J_0$ = $J(r(u, v_0))$ and otherwise $J  = J(r(u, v))$. As required, this metric reduces to the Minkowski metric $ds^2 = -du \, dv$ on the shell, $v_0 = v$, or for the vacuum solution $M=0$. These coordinates cover the entire spacetime and are free of coordinate singularities, unlike the Schwarzschild coordinates, \cref{eq:vac_metric_tr_coords}, which are singular at the horizon, or the asymptotically Minkowski coordinates, \cref{eq:vac_metric_asymp_coords}, which only extend up to the [outermost] horizon at $\bar{u} \rightarrow \infty$. 

The coordinate transformation from \cref{eq:vac_metric_tr_coords} to \cref{eq:vac_metric_null_coords} gives a set of first-order ordinary differential equations for the dilaton,
\begin{align}
    \d_u r &= -\frac{1}{2} \left(1 - \frac{2M}{J} \right) \left(1 - \frac{2M}{J_0}\right)^{-1}, \label{eq:vac_ODEs_u} \\
    \d_v r &= \frac{1}{2} \left(1 - \frac{2M}{J}\right) \label{eq:vac_ODEs_v}.
\end{align}

\section{Black Hole Models}\label{sec:black_hole_models}
The structure of the spacetime is entirely specified by the metric function $J(r)$. For the spacetime to be asymptotically flat the function $J(r)$ must grow at least as fast as $r$ as $r \rightarrow \infty$. The simplest such function is $J(r) = r$, in which case the classical theory is that of spherically symmetric gravity (SSG) and \cref{eq:vac_metric_tr_coords} is the Schwarzschild metric. 
The scalar curvature for the metric \cref{eq:vac_metric_tr_coords} is
\begin{equation}
    R = \frac{2M}{J} \left(2 \left( \frac{J^\prime}{J} \right)^2 - \frac{J^{\prime\prime}}{J} \right)
\end{equation}
which for $J = r$ reduces to 
\begin{equation}
    R = \frac{4M}{r^3},
\end{equation}
and there is a curvature singularity at $r = 0$. 

For a regular black hole spacetime the function $J(r)$ should satisfy the following properties: 
\begin{enumerate}
\item The spacetime should be asymptotically flat, which requires that $J(r)$ grows at least as fast as $r$ at large $r$. More specifically, for the metric to reproduce SSG far from the black hole requires $J(r) \sim r$ as $r \rightarrow \infty$;
\item The curvature is finite on $r > 0$. Generally this requires $J>0$;
\item The curvature remains finite as $r \rightarrow 0$. If at small $r$ the function $J$ behaves as $J(r) \sim r^n$ then the curvature $R = 2M r^{-(n+2)} (n^2 + n)$ diverges at $r=0$ if $n > -2$ and vanishes if $n < -2$. The most physically motivated choice is $n = -2$, in which case the curvature approaches a finite, constant value near $r = 0$ and the center of the black hole is replaced by a De Sitter core;
\item $J(r)$ has a single local minimum. This ensures that the black hole has at most two horizons for any mass. Although regular black holes can be constructed with any number of apparent horizons, we only consider two-horizon regular black holes.
\end{enumerate}

A family of metrics satisfying these conditions is given by functions of the form 
\begin{equation}\label{eq:reg_metric_family}
    J_a(r) = \frac{(r^a + l^a)^{3/a}}{r^2},
\end{equation}
$a>0$, which includes the Bardeen metric \cite{BardeenNonsingularGeneral1968} with $a = 2$ and a Hayward-like metric\footnote{In order to allow curvature-limited models \cite{FrolovNotesNonsingular2016} where the curvature does not diverge even with arbitrarily large mass, as in the Hayward model when written as $J = \frac{r^3 + M l^2}{r^2}$, $J$ would have to be a function of the mass function as well. This cannot be implemented consistently in this framework, but can be done, for example, as in \cite{KunstatterEtAlNew2D2016}.} \cite{HaywardFormationEvaporation2006} with $a = 3$. $l$ is a length parameter that determines the scale at which the metric transitions from Schwarzschild to De Sitter. The regularization scale is commonly taken as the Planck scale, so we will later set $l = l_p = 1$. The Bardeen metric function is shown in \cref{fig:metriccomp_bardeen_schwa}.

An apparent horizon is defined by the vanishing of the outgoing null expansion, $\partial_v r = 0$, which in the classical solution implies $J(r) = 2M$. The functions $J_a$ have a single minimum when $r^a = 2l^a$ and there are two horizons if $J(r_{\text{min}}) < 2M$ and no horizons if $J(r_{\text{min}}) > 2M$. At the critical mass 
\begin{equation}\label{eq:classical_Mcrit}
    M_{\text{crit}} = \frac{l}{2} \left( \frac{27}{4} \right)^{1/a}
\end{equation}
the black hole is extremal with a single horizon at ${r = 2^{1/a} l}$.

\begin{figure}[hpbt]
    \includegraphics[width=0.65\linewidth]{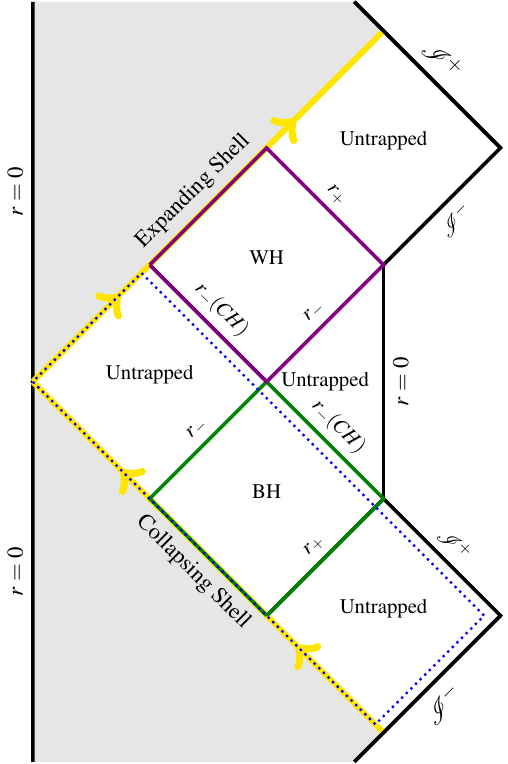}
    \caption{Penrose diagram for a classical non-singular black hole formed by the collapse of a null thin shell with reflective boundary conditions at $r=0$. The inner horizon of the black hole is a Cauchy horizon (CH). The blue dotted line shows the region that can be covered by our numerical simulations.}
    \label{fig:bardeen_cl_penrose}
\end{figure}

We will focus primarily on the Schwarzschild and Bardeen models; the functions $J_a$ for different values of $a$ are qualitatively similar to the latter. In the Schwarzschild model the dilaton field $r$ is directly related to the aerial radius in spherically symmetric gravity. In other models the connection to higher dimensional gravity is not as straightforward, but we consider the regularized spacetimes as corrections to the Schwarzschild spacetime and continue to interpret $r$ as a radial coordinate. Accordingly, the solution is limited to the region where $r \ge 0$. 

\begin{figure}[hpbt]
    \centering
    \includegraphics[width=\linewidth]{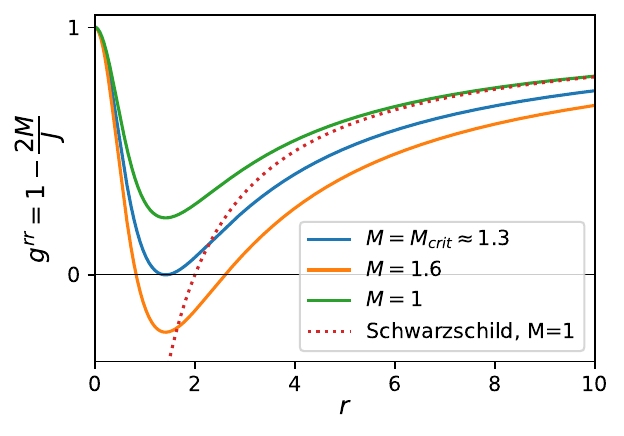}
    \caption{Components of the Bardeen metric with different masses ($l=1$). Horizons appear when $g^{rr} = 0$. The metric approximates the Schwarzschild metric at large $r$.}
    \label{fig:metriccomp_bardeen_schwa}
\end{figure}

\section{Hawking Radiation}\label{sec:hawking_radiation}
Hawking radiation is incorporated through the connection between Hawking radiation and the conformal trace anomaly in two dimensions \cite{ChristensenFullingTraceAnomalies1977}. The anomalous trace for a scalar field is $T = R/24$, which together with conservation of the stress-energy tensor, $\nabla_\mu T^{\mu \nu} = 0$, gives the components in general double-null coordinates 
\begin{equation}
    g_{uv} = -\frac{1}{2} e^{2\rho}, \quad g_{uu} = g_{vv} = 0
\end{equation}
as \cite{CallanEtAlEvanescentBlack1992}
\begin{subequations}\label{eq:anomolous_stresstensor_components}
\begin{align}
    T_{uv} &= -2\mu \, \d_u \d_v \rho \\
    T_{uu} &= 2\mu (\d_u \d_u \rho - \d_u \rho \, \d_u \rho + t_u(u))\label{eq:Tuu} \\
    T_{vv} &= 2\mu (\d_v \d_v \rho - \d_v \rho \, \d_v \rho + t_v(v))
\end{align}
\end{subequations}
where $\mu = N \hbar / 24$ is a parameter characterizing the strength of the quantum effects with $N$ scalar fields (henceforth we will work in units where $G = \hbar = 1$). The functions of integration $t_u(u), t_v(v)$ are fixed by imposing the boundary conditions that there is no radiation before the shockwave, $v \le v_0$, and no incoming radiation from past null infinity extept for the shell. For the coordinates described in \cref{sec:classical_dilaton_gravity} these boundary conditions require $t_u(u) = t_v(v) = 0$.
    
Solving the classical equations of motion, \cref{eq:EoMs_covariant_form_shell}, with this stress-energy tensor gives a semi-classical model of a collapsing black hole with Hawking radiation, including backreaction on the metric. In null coordinates the governing equations become
\begin{subequations}\label{eq:eom_null_coords}
\begin{equation}\label{eq:metric_eom_null_coords}
    J\d_u \d_v r + J^\prime \d_u r \, \d_v r + \frac{1}{4} e^{2\rho} J^\prime = -2\mu \d_u \d_v \rho 
\end{equation}
\begin{equation}\label{eq:dilaton_eom_null_coords}
    2 J \d_u \d_v \rho + J^{\prime\prime} \d_u r \, \d_v r + 2 J^\prime \d_u \d_v r + \frac{1}{4} e^{2\rho} J^{\prime\prime} = 0 
\end{equation}
\begin{equation}\label{eq:uu_contraint_null_coords}
    -J \d_u \d_u r + 2 J \d_u \rho \, \d_u r = 2\mu (\d_u \d_u \rho - \d_u \rho \, \d_u \rho) 
\end{equation} 
\begin{equation}\label{eq:vv_constraint_null_coords}
    -J \d_v \d_v r + 2 J \d_v \rho \, \d_v r = 2 \mu (\d_v \d_v \rho - \d_v \rho \d_v \rho).
\end{equation}
\end{subequations}
The first two equations are the dynamic equations of motion, and the last two equations are constraints that enforce the boundary conditions and are conserved by the dynamic equations.

It will later be convenient to write the first two equations in an alternate form,
\begin{subequations}\label{eq:eoms_form2}
    \begin{equation}\label{eq:eoms_form2_a}
        Q(r) \, \partial_u \partial_v r + P(r) \left(\partial_u r \, \partial_v r + \frac{1}{4} e^{2\rho} \right) = 0 ,
    \end{equation}
    \begin{equation}\label{eq:eoms_form2_b}
        2Q(r) \, \partial_u \partial_v \rho + \Pi(r) \left(\partial_u r \, \partial_v r + \frac{1}{4} e^{2\rho} \right) = 0,
    \end{equation}
\end{subequations}
or in coordinate free form as
\begin{subequations}\label{eq:eoms_form3}
    \begin{equation}\label{eq:eoms_form3_a}
        J Q \nabla^2 r - 2 P \mathcal{M} = 0
    \end{equation}
    \begin{equation}\label{eq:eoms_form3_b}
        J Q R + 2 \Pi \mathcal{M} = 0
    \end{equation}
\end{subequations}

where
\begin{equation}
    \begin{gathered}
    Q(r) = J - 2\mu \frac{J^\prime}{J}, \quad P(r) = J^\prime - \mu \frac{J^{\prime\prime}}{J}, \\ \quad \Pi(r) = J^{\prime\prime} - 2\frac{(J^\prime)^2}{J},
    \end{gathered}
\end{equation}
and 
\begin{equation}
    \mathcal{M} = \frac{J}{2} \left(1 - (\nabla r)^2 \right)
\end{equation}
is a generalized Misner-Sharp mass function.

\section{Semi-classical black holes}\label{sec:semi_classical_models}
Previously we discussed the conditions for the formation of trapped regions or singularities in the classical theory. We now consider the singularity and horizons in the semi-classical theory.

When radiation is introduced \cref{eq:eoms_form3} shows that a singularity is now present when $Q(r) = 0$, where the order of the equations changes, and the curvature must diverge because the roots of $\Pi$ or $\mathcal{M}$ do not generically coincide with the roots of $Q$. 

In the Schwarzschild model, $J(r) = r$, the singularity occurs when $r = 2\mu$. More generally, for any "Schwarzschild-like model", where $J$ has a zero and is monotonically increasing, the spacetime will always contain a singularity. On the other hand, if $J(r) > 0$ then it is always possible to choose a small, positive $\mu$ such that $\min Q(r) > 0$ and no singularity appears. Conversely, because $J^\prime$ must be positive at larger $r$ to satisfy the conditions described in \cref{sec:black_hole_models}, it is always possible to choose a large enough $\mu$ such that there will be a singularity, even in models that have no singularity classically. For the Bardeen model singularities appear if $\mu > \mu_{\text{crit}} \approx 6.56 l^2$. The breakdown occurs when the characteristic scale $\mu$ for radiation is large compared to the length scale of regularization $l$, and this singularity should more reasonably be viewed as a breakdown in the semi-classical approximation rather than a physical singularity.

In the semi-classical model no closed-form solution is known so the behavior of apparent horizons can only be found by solving the dynamic equations on the entire spacetime. However, because the solution joins smoothly to the interior flat solution, which is known, it can be determined whether a trapped region forms by examining the system on the shell. There the metric must match the flat interior, i.e. $\rho = 0$, and \cref{eq:vac_ODEs_u} implies $\d_u r = -\frac{1}{2}$. Defining 
\begin{equation}
    \chi(u) = \eval{\d_v r}_{v = v_0},
\end{equation}
\cref{eq:eoms_form2_a} reduces to an ordinary differential equation for $\chi$,
\begin{equation}\label{eq:chi_DE}
    Q \, \d_u \chi + P \left(-\frac{1}{2} \chi + \frac{1}{4} \right) = 0 \quad
\end{equation}
with solution 
\begin{equation}\label{eq:horizon_function}
    \chi(u) = -M (JQ)^{-\sfrac{1}{2}} + \frac{1}{2},
\end{equation}
with the integration constant fixed by requiring that the solution coincides with the classical solution at past null infinity. 

Horizons appear when $\chi = 0$. $\chi$ has a single local minimum when $P = 0$ and there are no horizons if the value of $\chi$ at the minimum is positive, and two horizons if this minimum is negative. The critical mass and the radius of the extremal horizon are found by solving the system \{$P(r, M; \mu) = 0, \chi(r, M; \mu) = 0$\} for $M$ and $r$. The critical mass decreases with larger radiation strength (\cref{fig:critical_mass}).

Since $Q$ is continuous it can be seen from \cref{eq:horizon_function} that if $Q$ has a root there will always be a solution to $\chi = 0$; that is, the critical mass becomes zero when $\mu \ge \mu_{\text{crit}}$, the smallest value of $\mu$ for which $Q = 0$ has a relevant solution. These singularities form inside the trapped region hidden by the apparent horizons, but can become naked as the black hole evaporates.

\section{Results: Schwarzschild}\label{sec:results_schwa}
We first briefly discuss the Schwarzschild model. In this case the classical action is the dimensional reduction of 4D general relativity with spherical symmetry. The classical and semi-classical solutions are shown in \cref{fig:spacetime_plot_schwa_classical,fig:spacetime_plot_schwa} respectively. In the classical spacetime the horizon is a null surface and meets the singularity at future null infinity. When radiation is included the horizon is a timelike surface that recedes to smaller $r$ and meets the singularity at finite time. 

\begin{figure}[hbpt]
    \centering
    \includegraphics[width=0.9\linewidth]{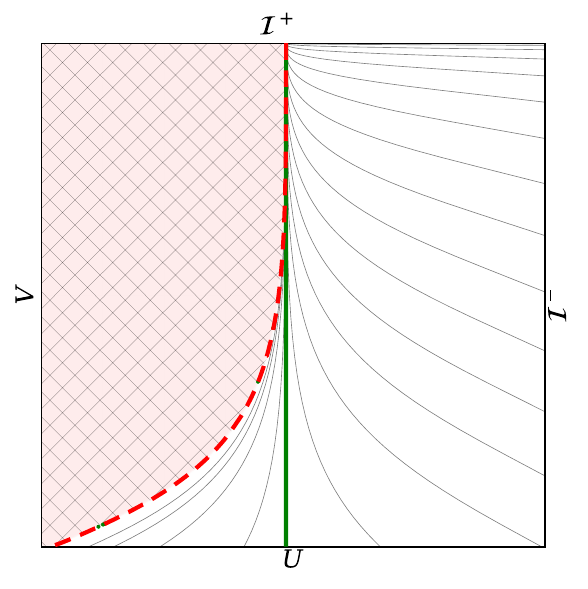}
    \caption{Classical Schwarzschild black hole with mass $M = 1$. Only the exterior solution is shown, with the shell lying along the bottom axis. The horizon is shown in green and the singularity in red. The light grey lines are curves of constant $r$.}
    \label{fig:spacetime_plot_schwa_classical}
\end{figure}

\begin{figure}[hbpt]
    \centering
    \includegraphics[width=0.9\linewidth]{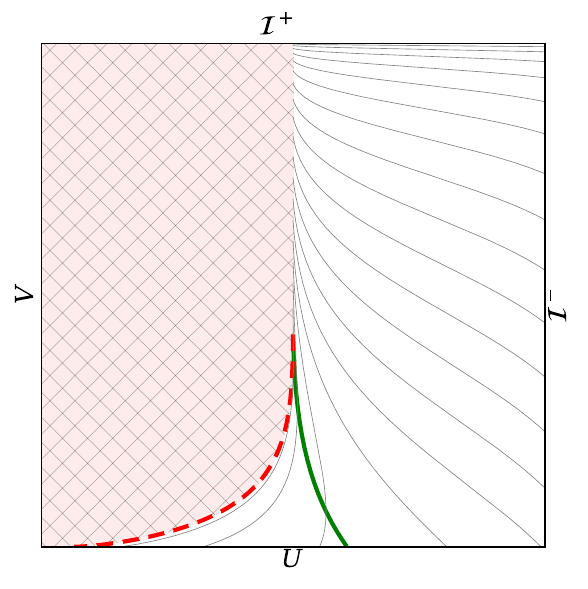}
    \caption{Evaporating Schwarzschild black hole, $M=1, \mu=1$.}
    \label{fig:spacetime_plot_schwa}
\end{figure}

Although the curvature diverges near the singularity, there is no "thunderbolt", or propagation of the curvature singularity, once it meets the horizon; after the trapped region vanishes, the curvature remains finite near the last ray and vanishes near $\scri^+$ (see \cref{fig:schwa_curvature}). This is in agreement with numerical simulations of other evaporating singular black holes \cite{AshtekarEtAlEvaporationTwodimensional2011}. Thus while the numerical calculation cannot continue past this meeting point, it could potentially be extended with suitable boundary conditions as in \cite{RussoEtAlEndPoint1992}, or by using suitable non-null coordinates such as Painleve-Gullstrand coordinates as in \cite{ZiprickKunstatterNumericalStudy2009}.
\begin{figure}[hpbt]
    \centering
    \includegraphics[width=\linewidth]{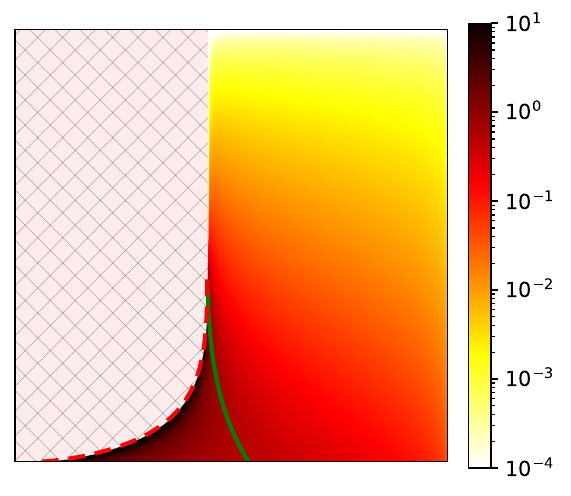}
    \caption{Scalar curvature in the evaporating Schwarzschild black hole.}
    \label{fig:schwa_curvature}
\end{figure}

\section{Results: Bardeen}\label{sec:results_bardeen}

We now focus on a regular black hole, in particular the Bardeen model \cite{BardeenNonsingularGeneral1968}, with metric function 
\begin{equation}
    J = \frac{(r^2 + 1)^{3/2}}{r^2}.
\end{equation} 
The numerical solutions for the classical and semi-classical Bardeen models are shown in \cref{fig:spacetime_plot_bardeen_classical} and \cref{fig:spacetime_plot_bardeen}. The simulations cover the region exterior to the collapsing shell up to the null line originating where the shell hits $r=0$. Because $r=0$ is a timelike curve part of the spacetime is therefore not reached. Appropriate boundary conditions for the matter field at $r=0$ would need to be imposed to extend the simulation to this region. A reasonable choice would be a reflection of the shell and matching to a new flat interior solution (see \cref{fig:bardeen_cl_penrose}).

For masses larger than the critical mass the solution now has two horizons but remains regular everywhere. In the semi-classical solution the horizons approach each other until meeting and annihilating. 

\begin{figure}[hbpt]
    \centering
    \includegraphics[width=0.9\linewidth]{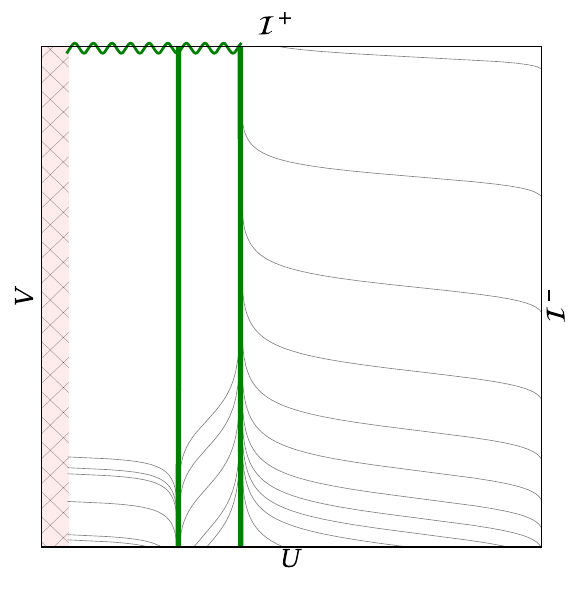}
    \caption{Classical Bardeen black hole with mass $M=1.33$. The rippled green line is the Cauchy horizon at $v \rightarrow \infty$.}
    \label{fig:spacetime_plot_bardeen_classical}
\end{figure} 
\begin{figure}[hbpt]
    \centering
    \includegraphics[width=0.9\linewidth]{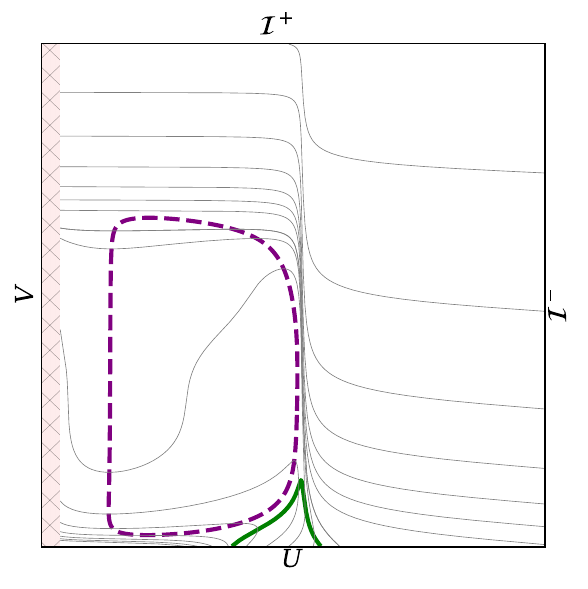}
    \caption{Evaporating Bardeen black hole with $M=1.33, \mu=0.5$.  The past horizon $\d_u r = 0$ is shown in purple.}
    \label{fig:spacetime_plot_bardeen}
\end{figure}

\begin{figure}[hbpt]
    \centering
    \includegraphics[width=0.9\linewidth]{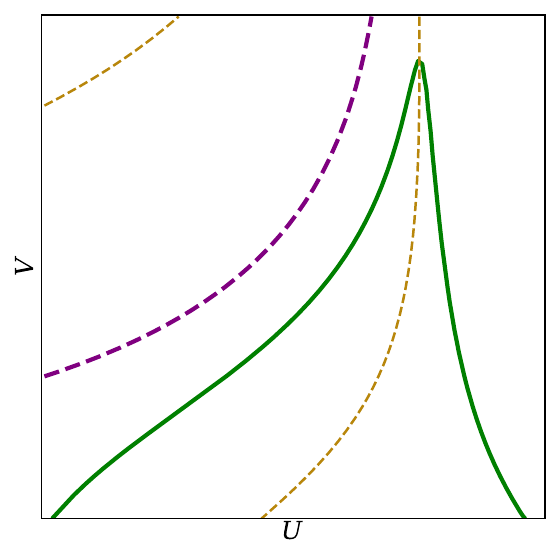}
    \caption{Zoom in on \cref{fig:spacetime_plot_bardeen} showing the trajectory of the apparent horizons. The yellow curves are $P(r)=0$.}
\end{figure}

The horizons meet smoothly with slope 
\begin{equation}
    \Delta :=-\frac{\d_u \d_v r}{\d_v \d_v r}
\end{equation}
going to zero as the two horizons approach each other (see \cref{fig:horizon_slope}). From \cref{eq:eoms_form2_a} this implies that the horizons meet at the radius satisfying $P(r) = 0$. Additionally, on the horizon $\partial_v r = 0$, so the mass function satisfies $\mathcal{M} = \frac{J(r)}{2}$. Therefore the horizons meet at a fixed radius and with mass function that is independent of the initial mass and depends only on the radiation parameter $\mu$.  

\begin{figure}[hpbt]
    \centering
    \includegraphics[width=\linewidth]{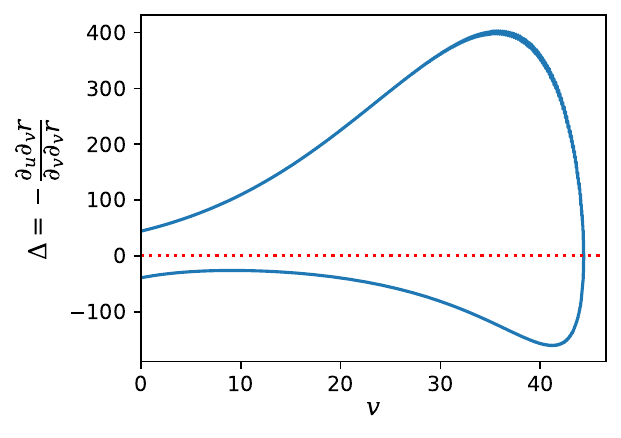}
    \caption{Slope of the apparent horizons.}
    \label{fig:horizon_slope}
\end{figure}

\begin{figure}[hpbt]
    \centering
    \includegraphics[width=\linewidth]{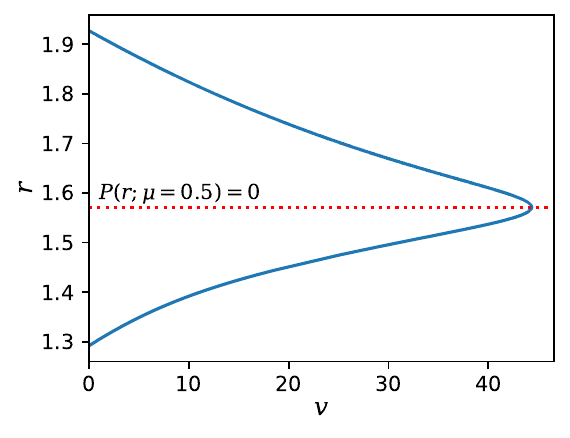}
    \caption{Value of the dilaton along the apparent horizons.}
\end{figure}

As mentioned in \cref{sec:semi_classical_models}, $P(r) = 0$ also defines the critical radius where the horizon forms on the shell for an extremal black hole. In the quasi-static approximation for Hawking radiation an extremal black hole has zero temperature and is eternal, and so it may be expected that the system settles into an extremal state, as has been suggested by previous investigations of regular black holes in 2D dilaton gravity \cite{LoweOLoughlinNonsingularBlack1993,DibaLoweNearextremalBlack2002,CadoniEtAlEvaporationInformation2023}. However, we find that this is not the case in our model. This is verified by calculating the horizon function $\partial_v r$ on a few slices $v > v_{\text{meet}}$ (see \cref{fig:horizon_function}). A minimum of 0 would indicate the presence of a single apparent horizon and an extremal black hole. For $v > v_{\text{meet}}$, $\partial_v r$ remains positive, approaching $1/2$ as $v \rightarrow \infty$, indicating that the spacetime lacks any horizons after the meeting point.

\begin{figure}[hpbt]
    \centering
    \includegraphics[width=\linewidth]{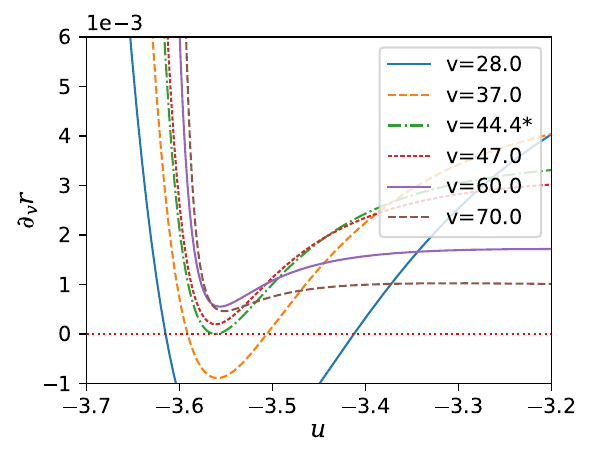}
    \caption{Horizon function $\partial_v r$ on selected slices of $v$. When a trapped region is present $\partial_v r$ crosses 0 twice, indicating the presence of two apparent horizons. The trapped region disappears when the apparent horizons meet at a single point ($v$ slice marked by an asterisk). At later $v$ the horizon function remains positive, signifying that there is no apparent horizon present.}
    \label{fig:horizon_function}
\end{figure}

In addition to the trapped region, an anti-trapped region where both null expansions are positive, ${\d_u r > 0, \d_v r > 0}$, also forms for large enough masses. The critical mass of the white hole is similar to the critical mass of the black hole, approaching the classical value $\sqrt{27/16}$ as $\mu \rightarrow 0$ and approaching 0 as $\mu \rightarrow \mu_{\text{crit}}$. 

The anti-trapped region appears to be generic in non-singular models. Similar structure is seen in other metrics of the form \cref{eq:reg_metric_family}, and evidence of an anti-trapped region was seen in previous work on other regular dilaton models, though not identified as such\footnote{See for example Fig 2 in \cite{DibaLoweNearextremalBlack2002}. The curve $\tilde{\phi}_0$ corresponds to $P(r) = 0$, and the "additional contour" suggests the presence of an anti-trapped region.}.

\begin{figure}[hbpt]
    \centering
    \includegraphics[width=\linewidth]{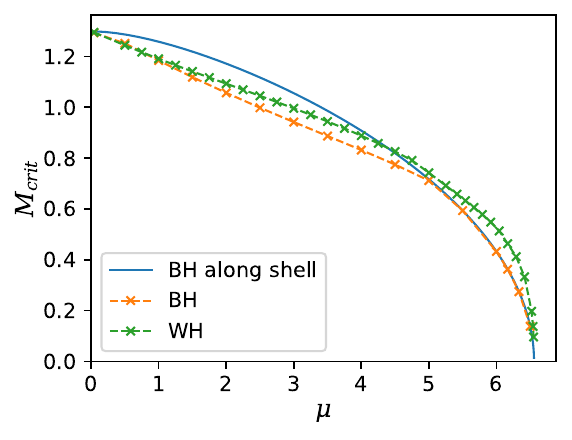}
    \caption{Critical mass for trapped (BH) and anti-trapped (WH) regions to form. Note that the critical mass for trapped regions to form along the shell can be found analytically, while the critical mass for anti-trapped regions or trapped regions that form off of the shell can only be found numerically.}
    \label{fig:critical_mass}
\end{figure}

\begin{figure}[hbpt]
    \centering
    \includegraphics[width=0.9\linewidth]{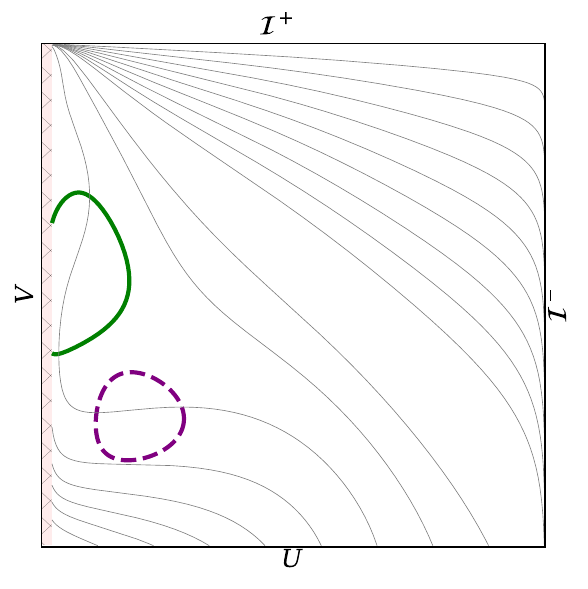}
    \includegraphics[width=0.9\linewidth]{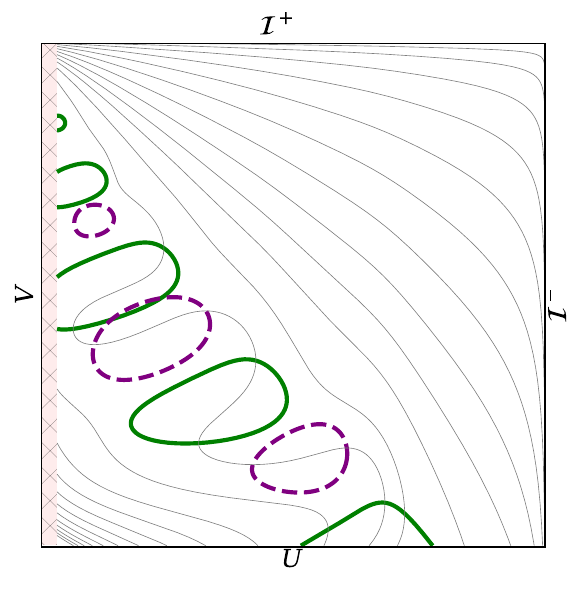}
    \caption{Plots of the Bardeen black hole for different model parameters. (Top) $M = 1.13, \mu=2$. Trapped and anti-trapped regions form dynamically even though no horizons appear on the shell. The white hole forms a closed region. (Bottom) $M=0.77, \mu=5.5$. Several closed trapped and anti-trapped regions form away from the shell.}
\end{figure}

For small masses the anti-trapped regions are compact. For masses above a critical mass (which depends on the value of $\mu$), the implicit solution (see \cref{sec:numerical_methods}) breaks down before it can resolve the endpoint of the white hole. The explicit method does not break down, and reveals that the anti-trapped region is compact. However, while a solution can be obtained, the solution in the upper left region of the spacetime does not converge in some cases because of a small region of large error where $\rho$ and $\d_u \rho$ grow large (this is the same region where the implicit solution fails), and the position at which the white hole closes is sensitive to the coordinate resolution and the time-stepping method used to solve the ODE. The qualitative behavior is consistent with the converged solutions, and so we conjecture that the anti-trapped region is indeed compact at the larger masses as well. However we can not rule out that that above a critical mass the white hole becomes eternal but is highly unstable to perturbations, here introduced by discretization error. If this is the case we expect it to be a feature of the infinitely thin shell approximation so that for finite shells the anti-trapped surface is compact and stable. 

At large $\mu$ the backreaction effects cause a series of trapped and anti-trapped regions to form. Additionally, it is possible for backreaction effects to cause \mbox{[anti-]} trapped regions to form dynamically when no black hole forms along the shell. This is best seen in \cref{fig:critical_mass} where the minimum mass required to form a black hole along the shell is larger than the mass required to form a trapped region anywhere in the spacetime at smaller $\mu$.

Nevertheless, the system eventually reaches a state that is asymptotically flat but filled with radiation. 
At $v \rightarrow \infty$, $\partial_v \rho$ vanishes as $\partial_v \rho \propto v^{-2}$, so we may define coordinates that are Minkowski at $\scri^+$ as 
\begin{gather}
    ds^2 = - e^{2\tilde{\rho}} d\tilde{u} \, dv \\
    \tilde{\rho} = \rho - \rho_\infty, \quad d\tilde{u} = e^{2\rho_\infty} du, \quad \rho_\infty = \lim_{v\rightarrow \infty} \rho.
\end{gather}

The components of the stress-energy tensor in these coordinates are shown in \cref{fig:bardeen_stresstensor}.  

\begin{figure}[hbpt]
    \centering
    \includegraphics[width=0.9\linewidth]{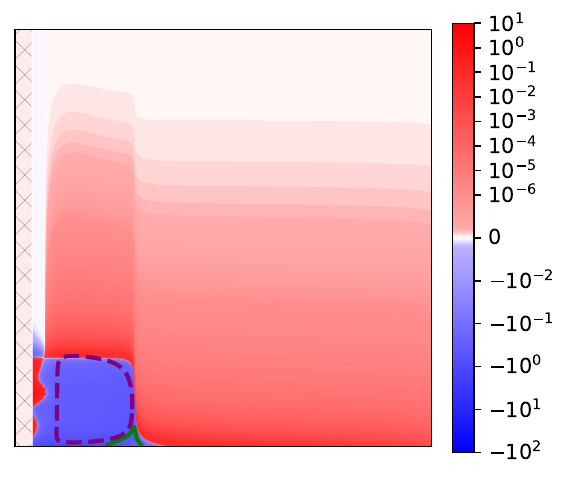}
    \caption{Curvature in the Bardeen model. The parameters for this plot are the same as in \cref{fig:spacetime_plot_bardeen} but with coordinates that cover more of the spacetime.}
    \label{fig:curvature_bardeen}
\end{figure}

\begin{figure*}[hpbt]
    \centering
    \includegraphics[width=\linewidth]{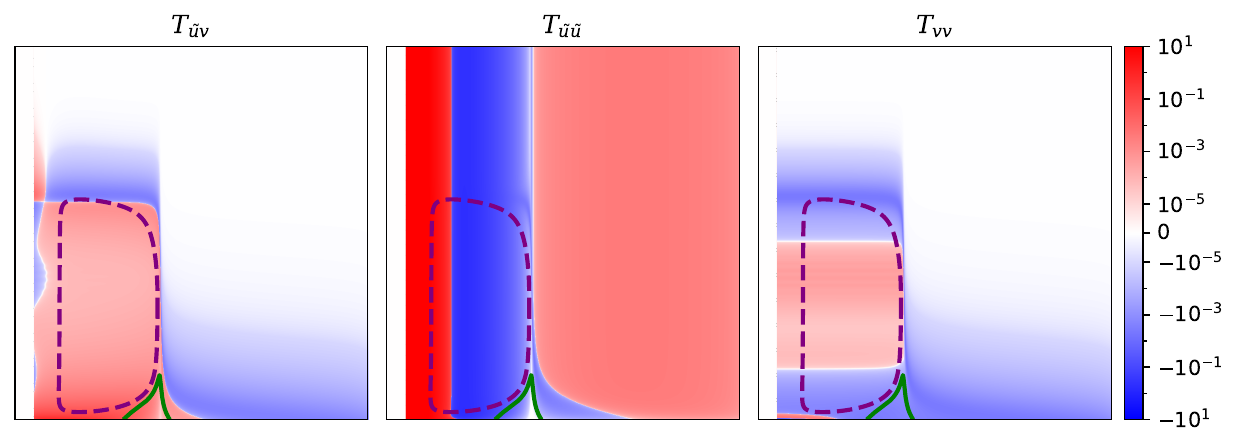}
    \caption{Components of the stress-energy tensor in asymptotic coordinates.}
    \label{fig:bardeen_stresstensor}
\end{figure*}

The black hole vanishing in finite time suggests that there is no Cauchy horizon and that the associated instabilities can be avoided. We now examine this in more detail. An outgoing null geodesic is given by
\begin{equation}
    n^\mu = \frac{dx^\mu}{d\lambda} = (0, e^{-2\rho}).
\end{equation}
The $v$ component can be inverted to find the affine parameter,
\begin{equation}
    \lambda = \int^v e^{2\rho} dv.
\end{equation}

In contrast to the classical solution, when radiation is present $e^{2\rho}$ remains finite, meaning that $\lambda$ is infinite as $v \rightarrow \infty$ and the spacetime is inextensible; that is, $v = \infty$ corresponds to a point on $\scri^+$ for all $u$, and there is no Cauchy horizon at $v = \infty$. 

Defining the frequency by the affine parameter \cite{FrolovInformationLoss2014}, 
\begin{equation}
    \omega = \frac{dv}{d\lambda},
\end{equation}
the geodesic equation implies the frequency of the outgoing ray satisfies
\begin{equation}
    \omega = \omega_0 e^{-2 \rho},
\end{equation}
where $\omega_0$ is the initial frequency. The vanishing of $e^{2\rho}$ in the classical solution leads to an infinite blueshift, while $e^{2\rho}$ remaining finite in the semi-classical model prevents the ray from being infinitely blueshifted. 
\begin{figure}[hpbt]
    \centering
    \includegraphics[width=\linewidth]{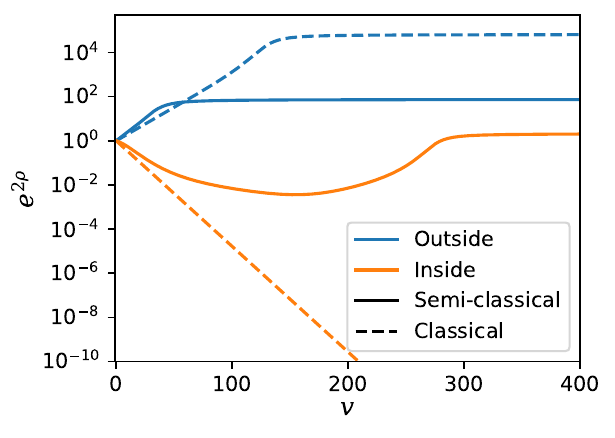}
    \caption{Conformal factor $e^{2\rho}$ evaluated on outgoing slices originating outside and inside the trapped region for the classical and semi-classical Bardeen spacetimes.}
    \label{fig:bardeen_conffactor}
\end{figure}

Similarly, mass inflation is avoided in the semi-classical model. As a simple model for mass inflation we consider the DTR relation, where perturbations are modeled as crossing ingoing and outgoing thin shells (see \cite{BarrabesEtAlCollisionLightlike1990,DiFilippoEtAlInnerHorizon2022,Carballo-RubioEtAlViabilityRegular2018} for more details). The effect of the perturbation on the final mass function behaves as 
\begin{equation}
    \Delta \mathcal{M} \propto \frac{-\mathcal{M}_{\text{in}}}{(\nabla r)^2}
\end{equation}
evaluated at the point the shells cross, where $\mathcal{M}_{\text{in}}$ is the mass of the ingoing shell, typically assumed to follow an inverse power law. As seen in \cref{fig:divrsq}, in the classical solution $(\nabla r)^2$ vanishes exponentially near the Cauchy horizon at $v \rightarrow \infty$, causing the mass to diverge. When radiation is added there is no Cauchy horizon present and $(\nabla r)^2$ remains finite at large $v$, and $\Delta \mathcal{M}$ vanishes rather than diverges. 

\begin{figure}[hpbt]
    \centering
    \includegraphics[width=\linewidth]{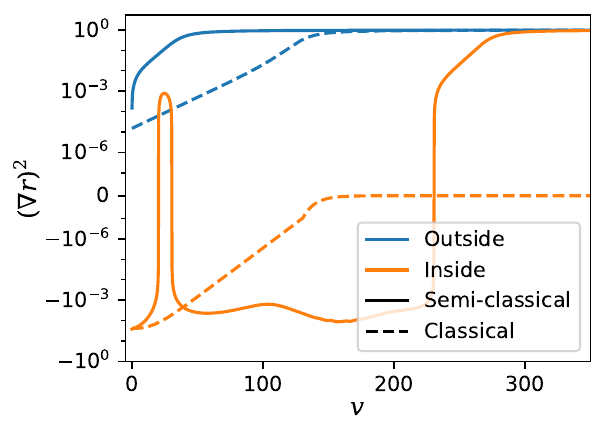}
    \caption{$(\nabla r)^2$ evaluated on outgoing slices. The exponential behavior near the classical inner horizon is suppressed by the y-axis scale, which is linear between $[-10^{-7}, 10^{-7}]$. Along the ray originating inside the trapped region of the semi-classical black hole, $(\nabla r)^2$ is negative inside the trapped region, becomes positive as it escapes the black hole, and is negative again as the ray passes through the anti-trapped region.}
    \label{fig:divrsq}
\end{figure}

\section{Discussion}\label{sec:discussion}
We find that a non-singular black hole evaporates smoothly in finite time and that the horizons disappear completely. This is consistent with other investigations of evaporating regular black holes \cite{LoweOLoughlinNonsingularBlack1993,DibaLoweNearextremalBlack2002,CadoniEtAlEvaporationInformation2023}. However, in contrast to previous results, we find that the trapped region disappears completely and the final spacetime is free of global horizons, as opposed to settling into an extremal state with an eternal horizon. Note that the methodology -- and in particular the initial condition -- in these works differs from those discussed in this paper; there, a static extremal black hole is perturbed by an incoming flux of mass, whereas in our case the black hole is formed by a shell collapsing into an initial vacuum state.

In addition to the black hole, we found that a white hole generically forms and that additional trapped and anti-trapped regions can appear and disappear as radiation is emitted.\footnote{This structure is similar to the time-symmetric bounce proposed for gravitational collapse (e.g. \cite{BarceloEtAlLifetimeProblem2015} and references therein).} Complex dynamics driven by the backreaction eventually settles, as all trapped and anti-trapped regions disappear and the system approaches an asymptotically flat spacetime. While this is shown conclusively for very small black holes, numerical issues limit the space of parameters that we can simulate and  it is difficult to verify for masses outside a small range. The apparently discrete change in dynamical structure is reminiscent of critical behavior, but further investigation is required to determine if larger black holes exhibit the same drama-free behavior as smaller black holes or if there is a phase transition to eternal black holes above a critical mass. This is currently in progress.

The absence of any singularities or trapping regions at late times implies that information can escape entirely to infinity. The devil, however, is in the details. For example, do the dynamics allow sufficient information to escape before the Page time \cite{PageInformationBlack1993}. Future work should study information in more detail. 

The absence of a Cauchy horizon has important implications for the possibility of non-singular black holes. While they resolve the central singularity, two-horizon solutions present their own problems. The Cauchy horizon represents a breakdown of predictability and further introduces instabilities as the mass \cite{PoissonIsraelInnerhorizonInstability1989} or stress-energy tensor \cite{HollandsEtAlQuantumInstability2020} catastrophically diverge under arbitrarily small perturbations, leading to new singularities.

Previous investigations raised the hope that Hawking radiation may alleviate the issue of mass inflation \cite{BonannoEtAlRegularEvaporating2023}, but while the inflation may be weakened in some evaporating models, evaporating regular black holes may still be subject to mass inflation \cite{Carballo-RubioEtAlMassInflation2024a}. A key limitation of many models is that they treat Hawking radiation quasi-statically, with a black hole taking an infinite amount of time to evaporate, no matter how small initially. The extension of the inner horizon to $v\rightarrow \infty$ is precisely what drives mass inflation, and may be absent in realistic models of evaporation. In at least one work \cite{Carballo-RubioEtAlMassInflation2024a} this condition has been weakened to adiabatic evaporation with a black hole that evaporates in finite time. While such an approximation should be accurate at early times and large masses, it would break down as the black hole evaporates and the adiabatic assumption would be violated before the trapped region disappears. Even so, this raises the possibility that mass inflation may build up for massive black holes before they evaporate, even if the horizons meet in finite time and there is no Cauchy horizon. Due to numerical limitations we were not able to study larger mass black holes and so cannot address this question. Improved numerical methods are needed to extend our results to larger black holes.  Curvature limiting models \cite{FrolovNotesNonsingular2016} could also be helpful in studying larger mass black holes, and may additionally provide more physically realistic descriptions of regular black holes than those used here.

As a final comment, we add that while we have primarily focused on the Bardeen black hole, the framework laid out in this paper can be used to study a vast array of black hole models. Preliminary investigations of other metrics in the family \cref{eq:reg_metric_family} show results consistent with those discussed for the Bardeen model. This may suggest that the qualitative behavior is generic in regular black hole models.

\begin{acknowledgments}
GK is very grateful to Jonathan Ziprick and Tim Taves for helpful conversations. Authors gratefully acknowledge that this research was supported in part by Discovery Grants number 2020-05346 (AF) and 2018-0409 (GK) from the Natural Sciences and Engineering Research Council of Canada.
\end{acknowledgments} 

\appendix

\section{Numerical Methods}\label{sec:numerical_methods}
We employ two different methods for solving the dynamic equations of motion, one based on the method used by Ramazanoglu et al \cite{RamazanogluPretoriusTwodimensionalQuantum2010} and one based on the method of lines, combined with various techniques for improving accuracy. 

Both methods were validated by checking the convergence of the error with increasing grid size for three different models: the classical Schwarzschild model (where an analytical solution can be found and used to verify the numerical solution), and radiating Bardeen models with $M=0.77, \mu=5.5$ and $M=1.33, \mu=0.5$.
In the first two cases, the implicit method exhibits the expected $E \sim h^2$ behavior for the error with step size $\sim h$. The error shows the expected convergence $E \sim h^4$ when run with Richardson extrapolation for two grids, but the convergence is slower than theoretically expected with higher order extrapolation. With 4 or 6 grids for Richardson extrapolation the error quickly reaches a minimum, and in some cases average error may even increase at larger grid size when using higher order extrapolation; however it may still be preferable to use the larger grid as it improves the accuracy and resolution in the regions of interest, specifically near the horizons and at late $V$. The method of lines was tested with different linear multistep methods (LMM) and yielded similar behavior. The two step Adams-Bashforth method showed the expected $E \sim h^2$ error convergence, and higher order LLMs reduced the error but with slower than expected convergence. For the small $\mu$ radiating Bardeen model the behavior is similar for most of the solution space, but the region near $u_{\text{meet}}$, large $v$ is more prone to error and convergence is slower.

In general, the implicit method proves more accurate than the method of lines. However, Newton's method sometimes fails to converge in the upper left region of the spacetime for larger masses, whereas the method of lines is able to continue beyond this point. We therefore use a series of solutions on refined meshes as described in \cref{sec:mesh_refinement}, with the implicit method used on the initial solutions to minimize error carried to subsequent grids, and the method of lines used on the final mesh to obtain a solution that can extend beyond this region. The field redefinition described in \cref{sec:field_redefinition} is used on the early iterations to minimize error near $\scri^-$, where the classical field is divergent and the quantum correction is small, but is not used on the final grid as this is not the case near the trapped region. When it is possible to find a solution on the entire grid with the implicit solver we opt to do so. These cases are also used to validate the method of lines, and we find that the solutions agree well.

\subsection{Implicit methods}
The coordinates are discretized into a uniform grid with spacing $\Delta u$ ($\Delta v$) in the $u$ ($v$) direction and a second-order short-centered finite difference approximation for the fields $q = r, \rho$,
\begin{subequations}
    \begin{gather}
    q^{i-\half}_{j-\half} \approx \frac{1}{4} (q^i_j + q^{i-1}_j + q^i_{j-1} + q^{i-1}_{j-1} ) \\
    \partial_u q^{i-\half}_{j-\half} \approx \frac{1}{2 \, \Delta u} (q^i_j + q^i_{j-1} - q^{i-1}_{j} - q^{i-1}_{j-1}) \\
    \partial_v q^{i-\half}_{j-\half} \approx \frac{1}{2 \, \Delta v} (q^i_j + q^{i-1}_j - q^i_{j-1} - q^{i-1}_{j-1} ) \\
    \partial_u \partial_v q^{i-\half}_{j-\half} \approx \frac{1}{\Delta u \, \Delta v} ( q^i_j - q^i_{j-1} - q^{i-1}_j + q^{i-1}_{j-1})
    \end{gather}
\end{subequations}
is substituted into the dynamical equations. Here $q^i_j$ means the value of the field $q$ at the point $(u_i, v_j)$, and $q^{i-1/2}_{j-1/2}$ is the value of $q$ in the center of a lattice square (see \cref{fig:implicit_schema}). This gives a pair of coupled non-linear algebraic equations for $r^i_j$ and $\rho^i_j$ in terms of the values of the fields at positions $(u_i, v_{j-1})$, $(u_{i-1}, v_j)$, and $(u_{i-1}, v_{j-1})$ which are solved using the Newton-Raphson method.
\begin{figure}[hpbt]
    \centering
    \includegraphics[width=0.5\linewidth]{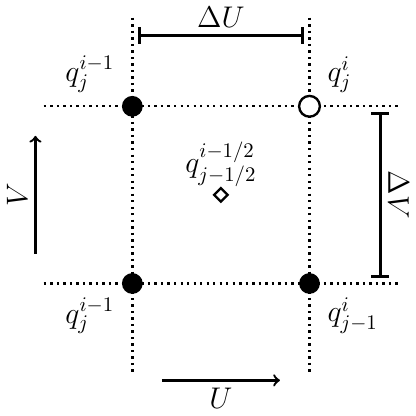}
    \caption{Schema of the implicit method. A finite difference approximation is made for the grid-centered point in terms of the neighboring grid points. At each step the values for the fields at $(u_i, v_{j-1})$, $(u_{i-1}, v_j)$, and $(u_{i-1}, v_{j-1})$ are known (the values at $i=0$ and $j=0$ are specified by boundary conditions) and we solve for $r$ and $\rho$ at $(u_i, j_i)$.}
    \label{fig:implicit_schema}
\end{figure} 

\subsection{Method of Lines}
Defining 
\begin{equation}\label{eq:MoL_v_derivs}
    \chi = \d_v r, \qquad \Sigma = \d_v \rho,
\end{equation}
\cref{eq:eoms_form2} are rewritten as a pair of first order ordinary differential equations,
\begin{subequations}
    \begin{equation}
        \d_u \chi = -\frac{P}{Q} \left(\d_u r \, \chi + \frac{1}{4} e^{2\rho} \right),
    \end{equation}
    \begin{equation}
        \d_u \Sigma = -\frac{\Pi}{2Q} \left(\d_u r \, \chi + \frac{1}{4} e^{2\rho} \right).
    \end{equation}
\end{subequations}
On a slice of constant $v$, given data for $r$ and $\rho$, these equations can be solved for $\chi$ and $\Sigma$, and \cref{eq:MoL_v_derivs} is then used to step forward $r$ and $\rho$ to the next $v$ slice. We use a predictor-corrector with a 4-step Adams-Bashforth method and 3-step Adams-Moulton method for stepping in both directions.

\subsection{Boundary conditions}
The spacetime should coincide with the classical solution on the shell and at past null infinity so the vacuum equations (\cref{eq:vac_ODEs_u,eq:vac_ODEs_v}) are used for the boundary conditions. The first equation reduces to ${\d_u r = -1/2}$ on the shell from which the boundary conditions ${r(u, v_0) = -u/2}$ are obtained. The second equation is evolved using the fourth-order Runge-Kutta method to obtain initial conditions $r(u_0, v)$. The conformal factor $\rho$ is 0 on the shell, and given by the vacuum metric, \cref{eq:vac_metric_null_coords}, on $u_0$.

The initial conditions on $u_0$ satisfy the constraints in the limit $u_0 \rightarrow -\infty$ by construction, but in practice the initial conditions are imposed at finite $u$, so there is some violation in the constraints; or equivalently, the initial conditions satisfy the constraints with a non-zero $t_v(v)$ term, which introduces incoming radiation from $\scri^-$. To minimize the impact of the incoming radiation, it is important that the initial conditions are imposed sufficiently far from the trapped region.

\subsection{Coordinate Compactification}
We introduce compactified coordinates to accomplish two goals. First, it maps the infinite spacetime onto a finite domain. Computationally the entire spacetime still cannot be captured because the coordinate transformation breaks down at infinity, requiring the coordinate grid to begin at some finite value, but the region that is covered can be made arbitrarily large. Second, the mapping increases resolution around the central position allowing better resolution near the center of the black hole where the dynamics evolve much faster. Any sigmoid function would accomplish these goals but we use the logistic function,
\begin{equation}
    U = \frac{e^{\kappa_u (u - u_c)}}{1 + e^{\kappa_u (u - u_c)}}, \qquad V = \frac{e^{\kappa_v (v - v_c)}}{1 + e^{\kappa_v (v - v_c)}}.
\end{equation}
The central parameters $u_c, v_c$ control the region where resolution is highest and the parameters $\kappa_u, \kappa_v$ control the scaling of the compactification. We fix $v_c = v_0 = 0$, and $u_c$ is chosen to maximize resolution in the region where the horizons meet. 

\subsection{Field Redefinition}\label{sec:field_redefinition}
The dilaton field $r$ and its derivatives diverge as ${u \rightarrow -\infty}$ or $v \rightarrow \infty$, making finite difference approximations quickly become inaccurate. However, at least part of this divergence is in the classical sector of the field as the semiclassical solution approaches the classical solution near $\scri^-$. Thus we can split the dilaton field into a classical part and a quantum correction, $r = r_c + \tilde{r}$, and replace the derivatives of the classical part with the expressions in \cref{eq:vac_ODEs_u,eq:vac_ODEs_v} to avoid taking finite differences of the divergent terms. In terms of the compactified coordinates and the new field the equations of motion become
\begin{widetext}
\begin{subequations}\label{eq:eoms_compact_scaled}
\begin{equation}
-J \frac{M \Jpv}{2 \Jv^2} e^{2\rho_c} + w J \d_U \d_V \tilde{r} + J^\prime \left( -\frac{1}{2} e^{2\rho_c} + w_u \d_U \tilde{r} \right) \left( \frac{1}{2} \left(1 - \frac{2M}{\Jv} \right) + w_v \d_V \tilde{r} \right) + \frac{1}{4} e^{2\rho} J^\prime + 2 w \mu \d_U \d _V \rho = 0
\end{equation}
\begin{equation}
    2wJ \d_U \d_V \rho + J^{\prime\prime} \left( - \frac{1}{2} e^{2\rho_c} + w_u \d_U \tilde{r} \right) \left( \frac{1}{2} \left(1 - \frac{2 M }{\Jv} \right) + w_v \d_V \tilde{r} \right) - J^\prime \frac{M \Jpv}{\Jv^2} e^{2\rho_c} + 2 w J^\prime \d_U \d_V \tilde{r} + \frac{1}{4} e^{2\rho} J^{\prime\prime} = 0
\end{equation}
\end{subequations}
\end{widetext}
where $w_u = \kappa_U (U-U^2)$, $w_v = \kappa_V (V-V^2)$, $w = w_u w_v$, $J_c = J(r_c)$, and $e^{2\rho_c}$ is the classical conformal factor given by \cref{eq:vac_metric_null_coords}. The boundary and initial conditions are taken as the classical solution, $\tilde{r} = 0$. 

A similar strategy is used when derivatives need to be calculated from the solution, in this case applying a similar procedure for $\rho$ as well.
 
% \subsection{Richardson Extrapolation}
% The finite difference approximation used in the implicit method is only second order accurate. However more accurate solutions can be obtained by combining simulations of different sizes. If $f_m$ is a second-order accurate approximation on a grid of size $mN_U \times mN_V$ then the combination
% \begin{equation}
%     \sum_{m=1}^p a_m f_m
% \end{equation}
% with the coefficients subject to the constraints
% \begin{equation}
% \begin{aligned}
%     & \sum_{m=1}^p a_m = 1 \\
%     & \sum_{m=1}^p \frac{a_m}{m^n} = 0, \qquad n = 2, 4, ..., 2(p-1)
% \end{aligned}
% \end{equation}
% will give a $2p$ order accurate approximation.

\subsection{Mesh refinement}\label{sec:mesh_refinement}
In order to maximize resolution in the region of interest the compactified coordinates should be chosen with a larger $\kappa_u$. However, this results in a grid that covers less of the physical spacetime, and the constraints are not satisfied if the classical solution is imposed on an initial slice close to the black hole. Therefore, to reduce the violation in the constraints at $\scri^-$ while achieving sufficient resolution around the black hole, simulations are first run on a coarse grid with small scaling parameter $\kappa_u$, so that the initial time slice is placed at a large (negative) value of physical coordinate $u$, and then on successively more zoomed-in grids with larger $\kappa_u$ to increase resolution in the dynamic region, using the data from the previous simulation as initial conditions. 

\bibliography{references}

\end{document}